\documentclass[conference]{IEEEtran}
\IEEEoverridecommandlockouts
\usepackage{cite}
\usepackage{amsmath,amssymb,amsfonts}
\usepackage{algorithmic}
\usepackage{graphicx}
\usepackage{textcomp}
\usepackage{xcolor}
\usepackage{tikz}
\usepackage{pgfplots}
\usepgfplotslibrary{groupplots}
\pgfplotsset{compat=1.12}
\usepackage{subfig}
\usepackage{algorithmic}
\usepackage{algorithm}
    \makeatletter 
\newcommand{\linebreakand}{%
  \end{@IEEEauthorhalign}
  \hfill\mbox{}\par
  \mbox{}\hfill\begin{@IEEEauthorhalign}
}
\makeatother 

\def\BibTeX{{\rm B\kern-.05em{\sc i\kern-.025em b}\kern-.08em
    T\kern-.1667em\lower.7ex\hbox{E}\kern-.125emX}}
\begin{document}

\title{Generative Adversarial Networks-Driven Cyber Threat Intelligence Detection Framework for Securing Internet of Things}

\author{\IEEEauthorblockN{
Mohamed Amine Ferrag}
\IEEEauthorblockA{\textit{Technology Innovation Institute} \\
\textit{9639 Masdar City, Abu Dhabi, UAE}\\
mohamed.ferrag@tii.ae}
\and
\IEEEauthorblockN{
Djallel Hamouda}
\IEEEauthorblockA{\textit{Labstic Laboratory} \\
\textit{Guelma University}\\
 Guelma, B.P. 401, 24000, Algeria \\
hamouda.djallel@univ-guelma.dz}
\and
\IEEEauthorblockN{
Merouane Debbah}
\IEEEauthorblockA{\textit{Technology Innovation Institute} \\
\textit{9639 Masdar City, Abu Dhabi, UAE}\\
merouane.debbah@tii.ae}
\and

\linebreakand 

\IEEEauthorblockN{
Leandros Maglaras}
\IEEEauthorblockA{\textit{Blockpass ID Lab} \\
\textit{Edinburgh Napier University}\\
Edinburgh EH10 5DT, UK \\
l.maglaras@napier.ac.uk}
\and
\IEEEauthorblockN{
Abderrahmane Lakas}
\IEEEauthorblockA{\textit{College of Information Technology} \\
\textit{United Arab Emirates University}\\
 Al-Ain, UAE \\
alakas@uaeu.ac.ae}
}

\maketitle

\begin{abstract}
While the benefits of 6G-enabled Internet of Things (IoT) are numerous, providing high-speed, low-latency communication that brings new opportunities for innovation and forms the foundation for continued growth in the IoT industry, it is also important to consider the security challenges and risks associated with the technology. In this paper, we propose a two-stage intrusion detection framework for securing IoTs, which is based on two detectors. In the first stage, we propose an adversarial training approach using generative adversarial networks (GAN) to help the first detector train on robust features by supplying it with adversarial examples as validation sets. Consequently, the classifier would perform very well against adversarial attacks. Then, we propose a deep learning (DL) model for the second detector to identify intrusions. We evaluated the proposed approach's efficiency in terms of detection accuracy and robustness against adversarial attacks. Experiment results with a new cyber security dataset demonstrate the effectiveness of the proposed methodology in detecting both intrusions and persistent adversarial examples with a weighted avg of 96\%, 95\%, 95 \%, and 95\% for precision, recall, f1-score, and accuracy, respectively.
\end{abstract}

\begin{IEEEkeywords}
IoT, Generative AI, GAN, Adversarial deep learning, Adversarial attacks.
\end{IEEEkeywords}

\section{Introduction}

The advent of the Internet of Things (IoT) has transformed our daily lives, work, and interactions with the environment. As IoT devices continue to evolve and demand more sophisticated features, the next iteration of wireless communication, 6G, has become a critical issue in the field of wireless technology. 6G is the next generation of wireless communication that will offer high-speed and low-latency connectivity to support a diverse range of IoT devices \cite{chafii2023twelve}. It is expected to offer even higher performance than 5G. It is expected to provide a bandwidth of up to 10 Gbps, a latency of 100 µs, and a data rate of up to 100 Gbps. Its energy efficiency is expected to be very high, which means that it can support a very large number of IoT devices without consuming excessive power. The network density for 6G IoT is expected to be 10 million or more devices per square kilometer. The architecture of 6G-enabled IoT devices will have a hierarchical structure that will consist of the following components: Device Layer, Network Layer, Application Layer, and Cloud Layer. The Device Layer encompasses the physical device and will house the necessary hardware and software components to facilitate 6G communication \cite{ferrag2023poisoning,ferrag2023generative}.

Our motivation is to improve the robustness of ML/DL-based cyber threat intelligence against adversarial evasion attacks. Several defense methods have been proposed in this context \cite{nicolae2018adversarial}, the most promising solution is adversarial training where the cyber threat intelligence model is trained on adversarial examples as well as the original examples (i.e real data) in order to make it more resilient to small perturbations in input data \cite{debicha2021adversarial}. However, it can lead to overfitting on the adversarial examples and decrease the generalization performance of the model. In addition, some generative methods may be more appropriate for certain types of data or models than others. {\color{black} Using GANs (Generative Adversarial Networks) to generate adversarial examples is one way to address these issues. GANs can generate more diverse and complex adversarial examples that are harder for the model to overfit on, compared to simpler methods like the Carlini-Wagner (CW) attack, DeepFool, Fast Gradient Sign Method (FGSM), etc.}

Our contributions to this paper are as follows :
\begin{itemize}
    \item We investigate the impact of FGSM adversarial attacks on the intrusion detection model.
    \item We propose a two-stage cyber threat intelligence using two detectors: the first detector uses GAN to detect adversarial examples, and the second detector is for intrusion detection. 
    \item  We evaluate the proposed GAN-based intrusion detection framework's performance in terms of detection accuracy as well as its resistance to adversarial evasion attacks.
\end{itemize}

\begin{figure}[h!]
\centering
\includegraphics[width=0.5\textwidth]{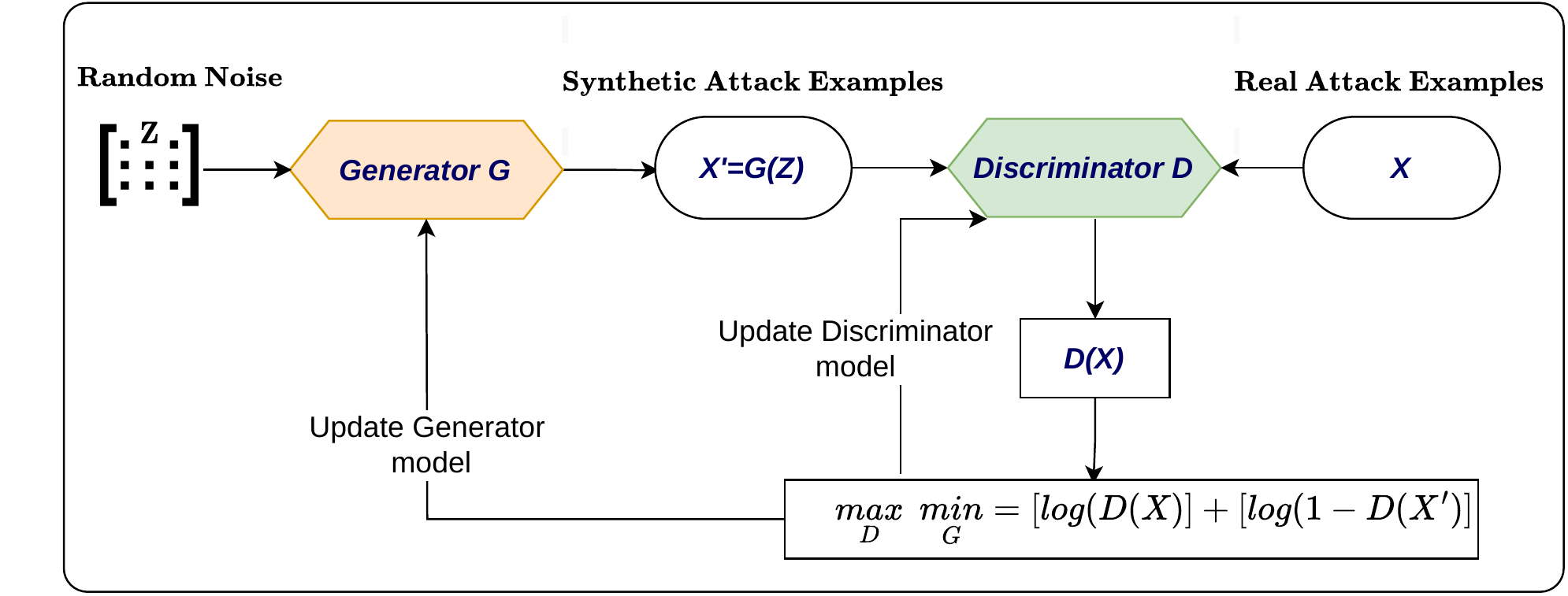}
\caption{The proposed GAN model training for adversarial examples detection} \label{fig1}
\end{figure}

\begin{algorithm}
\caption{Proposed methodology for GAN-based adversarial attack detection and intrusion detection}
\label{alg:1}
\begin{algorithmic}[1]
 \scriptsize
\REQUIRE Input data, pre-trained deep neural network, loss function, epsilon
\STATE Load the pre-trained deep neural network model
\STATE Choose a sample IoT data as input to the network
\STATE The gradient of the loss function is computed relative to the input IoT data using the following equation: 
\begin{equation}
\nabla_{x} J(\theta, x, y)
\end{equation}

\STATE Determine the sign of the gradient: 

\begin{equation}
sign(\nabla_{x} J(\theta, x, y))
\end{equation}

\STATE Multiply the sign of the gradient by a small constant $\epsilon$ to determine the magnitude of the perturbation:

\begin{equation}
\epsilon \cdot sign(\nabla_{x} J(\theta, x, y))
\end{equation}
\STATE Add the perturbation to the original IoT data to create an adversarial example: 

\begin{equation}
x_{adv} = x + \epsilon \cdot sign(\nabla_{x} J(\theta, x, y))
\end{equation}
{\color{black}
\STATE Feed the adversarial example into the deep neural network (CNN)
\STATE Compare the output of the network with the true label. If the output is incorrect, the attack was successful 
\STATE Use the first-stage detector, the GAN Discriminator model to detect generated successful adversarial attacks.
\STATE Use the second-stage detector, the CNN model to identify both normal and cyber attack types.}
\end{algorithmic}
\end{algorithm}

\begin{figure}[h!]
\centering
\includegraphics[width=0.5\textwidth]{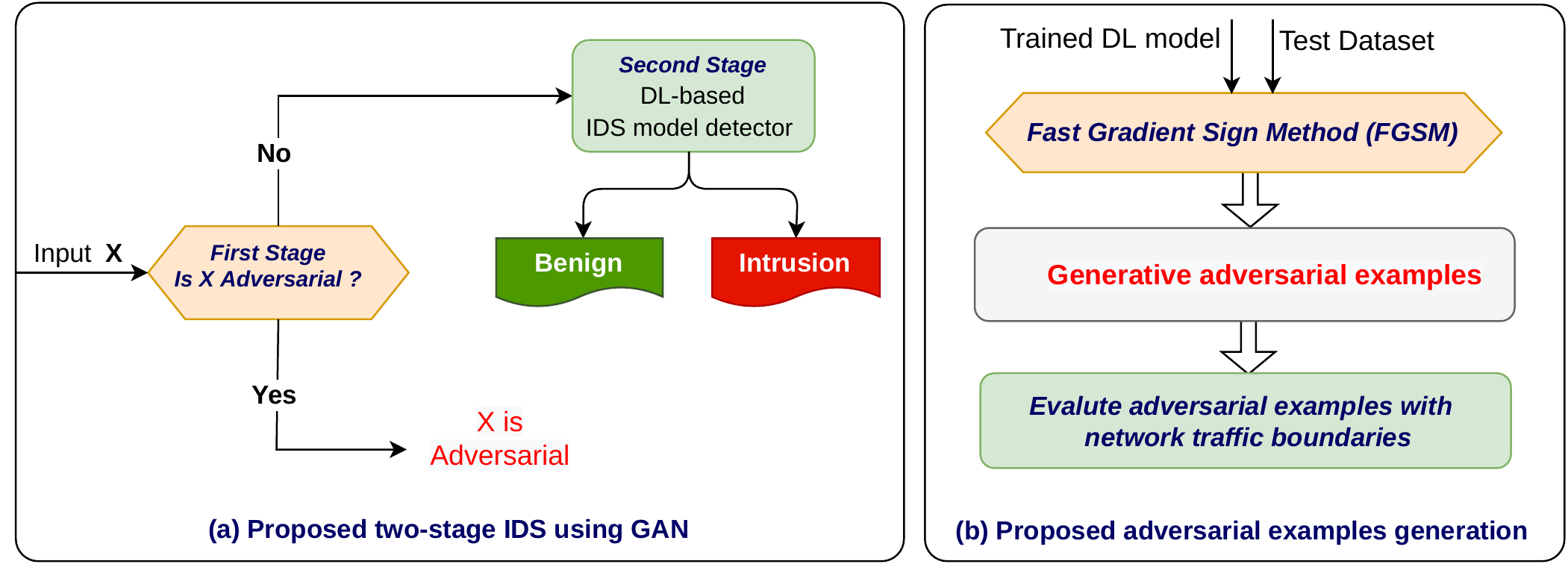}
\caption{A flow chart of the proposed cyber threat intelligence detection framework using two detectors.}\label{fig2}
\end{figure}

\section{Proposed methodology} \label{sec:3}
The proposed GAN-based model for detecting adversarial attacks is illustrated in Figure \ref{fig1}. To generate attack samples that mimic the real attack distribution, the generator is employed, while the discriminator is tasked with identifying genuine samples. Through dynamic minimum and maximum game theory, the generator and discriminator are trained, resulting in the production of artificial attack behaviors that closely resemble real attacks. The discriminator, on the other hand, can effectively differentiate between genuine and generated attacks.

Algorithm \ref{alg:1} and Figure \ref{fig2} present the structure of the proposed cyber threat intelligence framework using two detectors. First, input data is forwarded to the first GAN discriminator to detect evasion attacks before being forwarded to the DL classifier to identify real intrusions. GANs comprise two models, the generator and the discriminator. The generator creates new IoT data, while the discriminator determines if the generated IoT data is legitimate or not. The formula for the generator can be expressed as:
\begin{equation}
G(z) = f(z, \theta_g)
\end{equation}

Where $G(z)$ represents the generated IoT data, $z$ is the noise factor, and $\theta_g$ are the parameters for the generator. The formula for the discriminator can be depicted as:

\begin{equation}
D(x) = f(x, \theta_d)
\end{equation}

Where $D(x)$ stands for the discriminator's prediction, $x$ is the input data, and $\theta_d$ is the discriminator parameters.

To assess our detection approach, we utilize the advanced Fast Gradient Sign Method (FGSM) for creating adversarial instances \cite{goodfellow2014explaining}. This method does not require any knowledge of the target model's architecture or training data, which makes it a "model-agnostic" attack. As long as the attacker has access to the model's output for a given input, they can compute the gradients of the loss function and use FGSM to generate persistent adversarial examples. The aim is to decrease the highest level of distortion added to any characteristic that might result in misidentification. The FGSM procedure for IoT security is outlined in the following subsequent steps:

\begin{itemize}

 \item Step 1: Begin by loading the pre-trained deep neural network model.

 \item Step 2: Select a sample IoT data to be used as input for the network.

 \item Step 3: Compute the loss function gradient in relation to the input IoT data:

\begin{equation}
    \nabla_{x} J(\theta, x, y)
\end{equation}

where $J(\theta, x, y)$ is the loss function, $x$ is the input IoT data, and $y$ is the true label.

 \item Step 4: Determine the sign of the gradient:

\begin{equation}
sign(\nabla_{x} J(\theta, x, y))
\end{equation}

 \item Step 5: Multiply the sign of the gradient by a small constant $\epsilon$ to determine the magnitude of the perturbation:

 \begin{equation}
\epsilon \cdot sign(\nabla_{x} J(\theta, x, y))
\end{equation}

 \item Step 6: Add the perturbation to the original IoT data to create an adversarial example:

\begin{equation}
x_{adv} = x + \epsilon \cdot sign(\nabla_{x} J(\theta, x, y))
\end{equation}

 \item Step 7: Feed the adversarial example into the deep neural network and observe the output.

 \item Step 8: If the output doesn't match the true label, then the deep neural network is considered to be vulnerable to FGSM attacks, and the attack can be deemed successful.
\end{itemize}

The CNN model adopted by the Discriminator at first-stage detection for detecting adversarial attacks as well as at second-stage detection for detecting IoT attacks uses the following steps:
\begin{itemize}

\item Step 1: The initial stage of the CNN model involves the convolutional layer, which utilizes a collection of filters to obtain noteworthy features from the input IoT data. To perform this process, the convolution operation is applied as follows:

\begin{equation}
 (f * g)(n) = \sum_{m=-\infty}^{\infty}f(m)g(n-m)
\end{equation}

Where $f$ is the input IoT data. $g$ represents the filter, which is a smaller set of weights used to extract features from the input data. $n$ is the current index being evaluated in the output feature map. $m$ represents the index of the input data that is multiplied by the corresponding weight in the filter $g$. The summation over $m$ indicates that the filter is shifted across all possible positions in the input data to extract relevant features. The convolution operation calculates the dot product between the filter and the input data at every possible position, producing a feature map that represents the extracted features.

\item Step 2: Following the convolution operation, the subsequent step is to employ a non-linear activation function to the convolutional layer's output. In CNNs, the ReLU (rectified linear unit) activation function is widely used, which can be defined as:

\begin{equation}
f(x) = \max(0, x) 
\end{equation}

Where $x$ refers to the input value.

\item Step 3: we utilize the pooling layer to decrease the spatial dimensions of the feature maps, resulting in reduced computational complexity and the elimination of noise in the feature maps. To achieve this, we apply the max pooling technique, which can be defined as:

\begin{equation}
\max(x) = \max_{i=1}^{k} x_i
\end{equation}

Where $x$ is a set of values and $k$ is the size of the pooling window.

\item Step 4: To make predictions, the pooling layer's output undergoes processing via a multi-layer perceptron (MLP) network, also known as a fully connected layer. The formula for the fully connected layer is established as:

\begin{equation}
y = Wx + b 
\end{equation}

Where $W$ is the weight matrix, $x$ is the input IoT data vector, and $b$ is the bias vector.

\item Step 5: Following the fully connected layer, the output is subjected to a sigmoid or softmax activation function to produce final predictions for binary adversarial attack detection or multi-class attack detection, respectively.
\end{itemize}

\begin{figure} [h!]
\centering
\begin{tikzpicture}[scale=0.35]
\begin{groupplot}[group style={group name=myplot, group size=2 by 1}, height=8cm, width=12cm, scaled x ticks = false, ymin=0, ymax=1]
	
    \nextgroupplot[xmin=1,xmax=37, ymin=0.5, ymax=4, ylabel={$Loss\ \%$}, xlabel={Training checkpoints in 15 epochs }, legend style={legend pos = north east}, very thick, title={Generator loss},  ytick={.7,1,1.5,2,2.5,3.5}]

	 \addplot[color=red, mark=,] table[col sep=comma,header=false,x index=0, y index=2] {csvplot/GAN-IAM.csv};
	 \label{plots:Train-loss}
	    
    \nextgroupplot[xmin=1,xmax=37, ymin=0, ymax=.9, xlabel={Training checkpoints in 15 epochs },  xshift=0.5cm, legend style={legend pos = north east},very thick, title={Discriminator loss}, ytick={0,.1,.2,.3,.4,.5,.6,.7,.8} ]
    
    \addplot[color=blue, mark=,] table[col sep=comma,header=false,x index =0, y index=1] {csvplot/GAN-IAM.csv};

\end{groupplot}
\end{tikzpicture}
\caption{Comparison of the GAN models training loss at random checkpoints during training}
\label{fig:loss-results}
\end{figure}
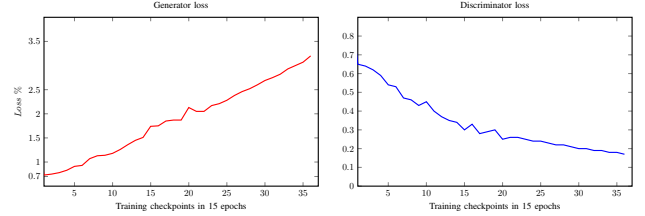

\begin{figure}[h!]
\centering
\includegraphics[width=0.15\textwidth]{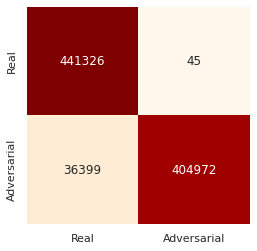}
\caption{Confusion matrix of Adversarial examples detection using GAN discriminator (first-stage detection)}
\label{fig:disc-confusion}
\end{figure}

\begin{table}[h!]
\centering
\caption{Evaluate the quality of generated adv-examples comparing with original data}
\tiny
\label{tab:adv_original_eval}
\begin{tabular}{|l|c|c|}
\hline
 & \textbf{Application features} & \textbf{Network features} \\ \hline
Mean perturbed features & 75.265282 & 17.610731 \\ \hline
Max perturbed features & 77 & 18 \\ \hline
Mean Euclidiant distance & 0.007527 & 0.001761 \\ \hline
Max Euclidiant distance & 0.0077 & 0.0018 \\ \hline
Mean Maximum perturbation & 0.009972 & 0.009973 \\ \hline
Max Maximum perturbation & 0.01 & 0.01 \\ \hline
\end{tabular}%
\end{table}

\begin{table}[h!]
\centering
\caption{Invalidity criteria of generated adversarial examples according to traffic boundaries }
\tiny
\label{tab:invalid-features}
\begin{tabular}{|l|c|c|}
\hline
\textbf{Criteria} & \textbf{Application  features} & \textbf{Network features} \\ \hline
Invalid value range & 99.71 \% & 99.72 \% \\ \hline
Invalid binary values & 99.56\% & 95 \% \\ \hline
Invalid class belonging & 99.7 \% & 0\\
\hline
\end{tabular}
\end{table}

\begin{table}[h!]
\setlength{\tabcolsep}{2.5pt}
\renewcommand{\arraystretch}{1}
\caption{Data distribution.}
\centering
\tiny
\begin{tabular}{|c|c|c|c|}
\hline
\textbf{Attack Classes} & \textbf{Train Count} & \textbf{Test Count} & \textbf{Total} \\ \hline
Normal & 1046926 & 323129 & 1370055 \\ \hline
Backdoor & 19890 & 4972 & 24862 \\ \hline
Vulnerability\_scanner & 40088 & 10022 & 50110 \\ \hline
DDoS\_ICMP & 93149 & 23287 & 116436 \\ \hline
Password & 40122 & 10031 & 50153 \\ \hline
Port\_Scanning & 18051 & 4513 & 22564 \\ \hline
DDoS\_UDP & 88027 & 22007 & 110034 \\ \hline
Uploading & 30107 & 7527 & 37634 \\ \hline
DDoS\_HTTP & 39929 & 9982 & 49911 \\ \hline
SQL\_injection & 40962 & 10241 & 51203 \\ \hline
Ransomware & 8740 & 2185 & 10925 \\ \hline
DDoS\_TCP & 40050 & 10012 & 50062 \\ \hline
XSS & 12732 & 3183 & 15915 \\ \hline
MITM & 320 & 80 & 400 \\ \hline
Fingerprinting & 801 & 200 & 1001 \\ \hline
\end{tabular}
\label{tab:tabd4}
\end{table}

\begin{table}[h!]
\setlength{\tabcolsep}{2.5pt}
\renewcommand{\arraystretch}{1}
\caption{The parameters of the CNN model used by the discriminator.}
\centering
\tiny
\begin{tabular}{|p{0.6in}|p{0.3in}|p{2in}|}
\hline
\textbf{Layer} & \textbf{Type} & \textbf{Parameters} \\ \hline
Convolutional 1 & Conv1d & Input Channels: 1, Output Channels: 64, Kernel Size: 3, Padding: 1 \\ \hline
Convolutional 2 & Conv1d & Input Channels: 64, Output Channels: 32, Kernel Size: 3, Padding: 1 \\ \hline
Convolutional 3 & Conv1d & Input Channels: 32, Output Channels: 16, Kernel Size: 3, Padding: 1 \\ \hline
Fully Connected 1 & Linear & Input Units: 95 * 16, Output Units: 30 \\ \hline
Output & Linear & Input Units: 30, Output Units: 15 \\ \hline
\end{tabular}
\label{tab:tab1}
\end{table}

\begin{table}[h!]
\setlength{\tabcolsep}{2.5pt}
\renewcommand{\arraystretch}{1}
\caption{Classification report for multi-classification of the discriminator.}
\centering
\tiny
\begin{tabular}{|c|c|c|c|c|}
\hline
\tiny
\textbf{Class}  & \multicolumn{1}{c|}{\textbf{Precision}} & \multicolumn{1}{c|}{\textbf{Recall}} & \textbf{F1-Score} & \textbf{Support} \\ \hline\hline
Normal & 1.00 & 1.00 & 1.00 & 323129\\ \hline
Backdoor & 0.96 & 0.94 & 0.95 & 4972 \\ \hline
Vulnerability\_scanner & 0.97 & 0.85 & 0.91 & 10022 \\ \hline
DDoS\_ICMP & 1.00 & 1.00 & 1.00 & 23287 \\ \hline
Password & 0.45 & 0.87 & 0.59 & 10031\\ \hline
Port\_Scanning & 0.92 & 0.52 & 0.67 & 4513 \\ \hline
DDoS\_UDP & 1.00 & 1.00 & 1.00 & 22007 \\ \hline
Uploading & 0.67 & 0.48 & 0.56 & 7527 \\ \hline
DDoS\_HTTP & 0.76 & 0.94 & 0.84 & 9982 \\ \hline
SQL\_injection & 0.76 & 0.23 & 0.35 & 10241 \\ \hline
Ransomware & 0.87 & 0.84 & 0.85 & 2185\\ \hline
DDoS\_TCP & 0.82 & 1.00 & 0.90 & 10012\\ \hline
XSS & 0.59 & 0.40 & 0.48 & 3183\\ \hline
MITM & 1.00 & 1.00 & 1.00 & 80\\ \hline
Fingerprinting & 0.89 & 0.54 & 0.67 & 200 \\ \hline \hline
\textbf{Accuracy} & & & \textbf{0.95} & 441371\\ \hline
\textbf{Macro avg} & 0.84 & 0.77 & 0.78 & 441371\\ \hline
\textbf{Weighted avg} & \textbf{0.96} & \textbf{0.95} & \textbf{0.95} & 441371\\ \hline
\hline
\end{tabular}
\label{tab:tabres2}
\end{table}

\begin{figure}[h!]
\centering
\includegraphics[width=0.4\textwidth]{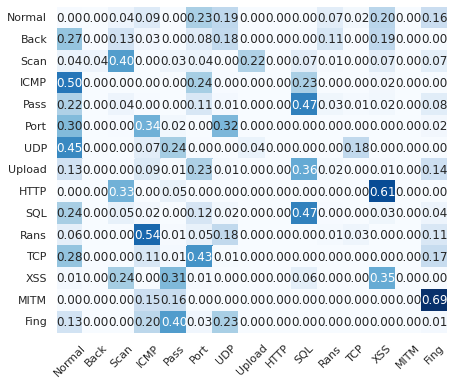}
\caption{Confusion matrix of CNN results against  adversarial attacks generated using FGSM with epsilon=0.01}\label{fig:adv-confu}
\end{figure}

\begin{figure}[t!]
\centering
\includegraphics[width=0.4\textwidth]{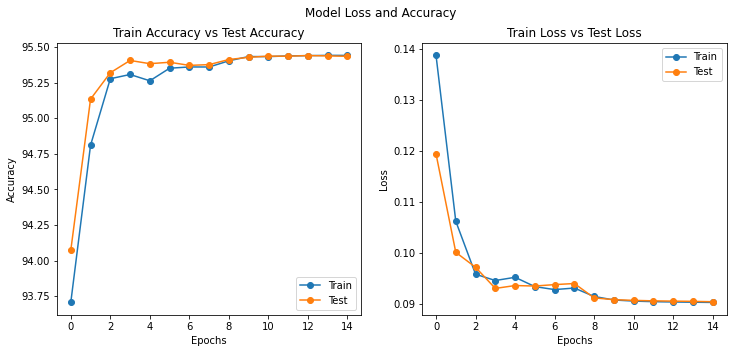}
\caption{Loss and Accuracy of CNN training and evaluation}
\label{fig:cnn-acc-loss}
\end{figure}

\begin{figure}[h!]
\centering
\includegraphics[width=0.4\textwidth]{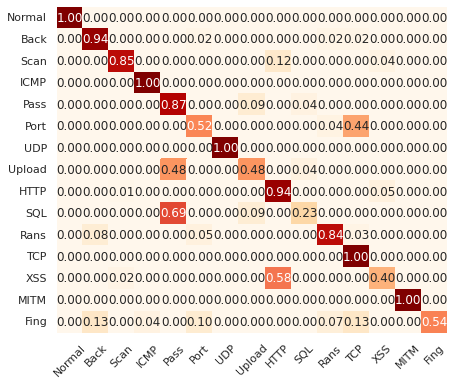}
\caption{Confusion matrix of proposed Cyber Threat Intelligence detection framework}
\label{fig:original-confu}
\end{figure}

\section{Performance Evaluation} \label{sec:4}
The proposed two-stage intrusion detection experiment results were performed on a Google Collaboratory-free environment using the PyTorch library. We used the Edge-IIoTset dataset \cite{ferrag2022edge} a recently proposed comprehensive and realistic cyber security dataset designed for IoT and IIoT applications. The dataset was constructed using a specialized IoT/IIoT testbed with various devices, sensors, and settings and includes over 10 types of IoT devices and 14 analyzed attacks that can be classified into 5 different threats. A total of 61 new features were extracted from the dataset's 1176 features and have high correlations.
{\color{black}
In order to assess the robustness of the proposed GAN model against evasion attacks, we first processed the  network traffic dataset,  cleaned duplicate and corrupted samples, performed one-hot encoding to categorical features, performed features scaling, and finally performed an analysis of network traffic criteria such as binary values, value ranges, and class belonging for categorical features.}
A series of adversarial examples were generated using the Fast Gradient Sign Method (FGSM) was employed to produce the adversarial validation set. The generated data was carefully selected using the minimum perturbation ($epsilon = 0.01$) to reflect network traffic boundaries and then evaluated using the GAN discriminator. {\color{black} The results of evaluating the quality of generated adversarial examples, grouped by application layer features and network traffic features, are shown in Table \ref{tab:adv_original_eval}. The results demonstrate that the generated adversarial examples are similar to the original data in terms of Euclidean distance. The number of perturbed features and maximum perturbation indicates that the generated attack examples were of high quality, with low levels of perturbation and small differences between the original and generated data, but there is a relatively high percentage of invalid data, particularly for application layer features. This may have implications for the practicality of evading attacks. However, these attacks can largely affect the accuracy and reliability of the models.  

Table \ref{tab:tabd4} presents the distribution of different types of attack classes in a train and test data set. The train data distribution table has 14 different attack classes and their respective counts, with a total of 1046926 normal instances and the rest being various types of cyber attacks. The test data distribution table has the same 14 attack classes with their respective counts, with a total of 323129 normal instances and the rest being various types of cyber attacks

Table \ref{tab:tab1} presents the parameters of the Convolutional Neural Networks (CNN) model used by the discriminator. The architecture consists of three convolutional layers, one fully connected layer, and an output layer. The convolutional layers use the ReLU activation function, which has been shown to be effective in many deep-learning applications. The first layer is a 1-dimensional convolutional layer with 64 output channels, a kernel size of 3, and padding of 1. The second and third layers are similarly designed with 32 and 16 output channels, respectively. The output of the third convolutional layer is passed through a max-pooling layer to reduce the spatial resolution. After the convolutional and pooling layers, the feature map is flattened and passed through a fully connected layer with 30 output units and ReLU activation. The final layer is a linear layer with 15 output units, and the output is passed through a log-softmax activation function to produce class probabilities.

\subsection{Results}

In order to demonstrate the impact of adversarial evasion attacks, we first trained a CNN classifier to use for FGSM attack generation. Figure \ref{fig:cnn-acc-loss} presents the Loss and Accuracy of CNN training and evaluation. The training and testing accuracies are reported for 15 epochs. The training accuracy starts at 93.707\% and increases to 95.442\% at the end of 15 epochs. The testing accuracy also starts at 94.073\% and increases to 95.435\% at the end of 15 epochs. These results indicate that the model is able to generalize well on unseen data. The decreasing trend in loss values indicates that the model is learning the underlying patterns in the data and improving its performance over time. The model performs well and can be considered a good solution for cyber threat detection. However, Figure \label{fig:adv-confu} demonstrates the impact of adversarial threats on a well-trained CNN classifier, where the model's accuracy dropped from 95.44\% to 2.55\%. We can see that the normal class is identified as malicious traffic, while the attack classes were largely identified as legitimate traffic.

Using our proposed first-stage detection strategy, Figure \ref{fig:loss-results} depicts the training loss of both GAN models, which indicates the discriminator's predictions compared to the input examples' ground truth reality. In the beginning, the discriminator has a high error rate (i.e., loss) and starts decreasing through training. Unlike the generator's loss, which begins low and rapidly increases. We can demonstrate that the discriminator has beaten the generator and has efficiently learned the representation of real input data. We further evaluate the classification performance using the confusion matrix, which is a representation of the true label versus the predicted label. Figure \ref{fig:disc-confusion} depicts the obtained results. The GAN discriminator was able to identify FGSM adversarial attacks with a recall of 96\% and real data with a recall of 100\%. These results demonstrate that the proposed GAN method was efficient in detecting high-quality adversarial threats.

Table \ref{tab:tabres2} presents the classification report for the multi-classification of the discriminator. The precision column shows the accuracy of the positive predictions made by the algorithm. The confusion matrix of the proposed cyber threat intelligence detection framework is presented in Figure \ref{fig:original-confu}. For example, the precision for the Normal category is 1.00, which means that all the instances classified as Normal by the algorithm were actually Normal. The Generative Adversarial Networks-Driven Cyber Threat Intelligence Detection Framework has demonstrated impressive results in classifying different types of cyber threats with a high level of accuracy. The model achieved an overall accuracy of 95\%, correctly identifying 419,302 out of 441,371 instances. The model showed a perfect precision and recall score for Normal activity and DDoS\_ICMP attacks, which had support values of 323,129 and 23,287, respectively. However, for some types of attacks, such as SQL\_injection and Port\_Scanning, the model showed lower recall scores of 0.23 and 0.52, respectively. Nevertheless, the model's overall performance was excellent, with a weighted average precision and recall score of 0.96 and 0.95, respectively. These results suggest that the Generative Adversarial Networks-Driven Cyber Threat Intelligence Detection Framework has great potential in identifying and preventing various types of cyber threats, making it a valuable tool for cyber security professionals.

\section{Conclusion} \label{sec:5}
{\color{black}
In this paper, we proposed a two-stage intrusion detection framework by employing generative adversarial networks (GANs). Specifically, we introduced a GAN model to improve robustness against adversarial attacks and a DL-based intrusion detection approach. We demonstrated the effectiveness of these methods in detecting persistent adversarial examples, generated using the FGSM method. In real-world scenarios, these adversarial examples may be generated intentionally or unintentionally as a result of software or hardware errors, resulting in poor cyber threat intelligence performance.}

\bibliographystyle{IEEEtran}
\bibliography{bibliography} 

\begin{thebibliography}{1}
\providecommand{\url}[1]{#1}
\csname url@samestyle\endcsname
\providecommand{\newblock}{\relax}
\providecommand{\bibinfo}[2]{#2}
\providecommand{\BIBentrySTDinterwordspacing}{\spaceskip=0pt\relax}
\providecommand{\BIBentryALTinterwordstretchfactor}{4}
\providecommand{\BIBentryALTinterwordspacing}{\spaceskip=\fontdimen2\font plus
\BIBentryALTinterwordstretchfactor\fontdimen3\font minus
  \fontdimen4\font\relax}
\providecommand{\BIBforeignlanguage}[2]{{%
\expandafter\ifx\csname l@#1\endcsname\relax
\typeout{** WARNING: IEEEtran.bst: No hyphenation pattern has been}%
\typeout{** loaded for the language `#1'. Using the pattern for}%
\typeout{** the default language instead.}%
\else
\language=\csname l@#1\endcsname
\fi
#2}}
\providecommand{\BIBdecl}{\relax}
\BIBdecl

\bibitem{chafii2023twelve}
M.~Chafii, L.~Bariah, S.~Muhaidat, and M.~Debbah, ``Twelve scientific
  challenges for 6g: Rethinking the foundations of communications theory,''
  \emph{IEEE Communications Surveys \& Tutorials}, 2023.

\bibitem{ferrag2023poisoning}
M.~A. Ferrag, B.~Kantarci, L.~C. Cordeiro, M.~Debbah, and K.-K.~R. Choo,
  ``Poisoning attacks in federated edge learning for digital twin 6g-enabled
  iots: An anticipatory study,'' \emph{arXiv preprint arXiv:2303.11745}, 2023.

\bibitem{ferrag2023generative}
M.~A. Ferrag, M.~Debbah, and M.~Al-Hawawreh, ``Generative ai for cyber
  threat-hunting in 6g-enabled iot networks,'' \emph{arXiv preprint
  arXiv:2303.11751}, 2023.

\bibitem{nicolae2018adversarial}
M.-I. Nicolae, M.~Sinn, M.~N. Tran, B.~Buesser, A.~Rawat, M.~Wistuba,
  V.~Zantedeschi, N.~Baracaldo, B.~Chen, H.~Ludwig \emph{et~al.}, ``Adversarial
  robustness toolbox v1. 0.0,'' \emph{arXiv preprint arXiv:1807.01069}, 2018.

\bibitem{debicha2021adversarial}
I.~Debicha, T.~Debatty, J.-M. Dricot, and W.~Mees, ``Adversarial training for
  deep learning-based intrusion detection systems,'' \emph{arXiv preprint
  arXiv:2104.09852}, 2021.

\bibitem{goodfellow2014explaining}
I.~J. Goodfellow, J.~Shlens, and C.~Szegedy, ``Explaining and harnessing
  adversarial examples,'' \emph{arXiv preprint arXiv:1412.6572}, 2014.

\bibitem{ferrag2022edge}
M.~A. Ferrag, O.~Friha, D.~Hamouda, L.~Maglaras, and H.~Janicke,
  ``Edge-iiotset: A new comprehensive realistic cyber security dataset of iot
  and iiot applications for centralized and federated learning,'' \emph{IEEE
  Access}, vol.~10, pp. 40\,281--40\,306, 2022.

\end{thebibliography}
\end{document}